\documentclass[lettersize,journal]{IEEEtran}
\usepackage{amsmath,amsfonts}
\usepackage{algorithmic}
\usepackage{algorithm}
\usepackage{array}
\usepackage[caption=false,font=normalsize,labelfont=sf,textfont=sf]{subfig}
\usepackage{textcomp}
\usepackage{stfloats}
\usepackage{url}
\usepackage{verbatim}
\usepackage{graphicx}
\usepackage{cite}
\usepackage{subcaption}
\usepackage{amsmath}
\usepackage{amssymb}
\usepackage{multirow}
\usepackage{lineno}
\usepackage[abs]{overpic} 
\begin{document}

\title{Raman-Accelerated Power Depletion of Fundamental Mode in a Few-Mode Fiber in the Visible Spectral Range}

\author{Wasyhun A. Gemechu*, Guohao Fu, and Mario Zitelli
\thanks{Manuscript received August x, 2025; revised Y Z, 2025.(\it{Corresponding author: Wasyhun A. Gemechu})}        
\thanks{Wasyhun A. Gemechu is with the Department of Information Engineering, Università degli Studi di Brescia, 25123 Brescia, Italy (e-mail: wasyhun.gemechu@unibs.it)}
\thanks{Guohao Fu is with the Department of Precision Instrument, Tsinghua University, Beijing 100084, China (e-mail: fgh21@mails.tsinghua.edu.cn)}
\thanks{Mario Zitelli is with the Department of Information Engineering, Electronics, and Telecommunications, Sapienza University of Rome, Via Eudossiana 18, 00184, Rome, Italy (e-mail: mario.zitelli@uniroma1.it)}
}
\markboth{Journal of \LaTeX\ Class Files,~Vol.~14, No.~8, August~2025}%
{Shell \MakeLowercase{\textit{et al.}}: A Sample Article Using IEEEtran.cls for IEEE Journals}


\maketitle

\begin{abstract}
We experimentally and numerically investigate Raman-driven power depletion in the fundamental mode of few-mode fibers (FMFs) excited by visible ultrashort pulses. Using a tunable femtosecond laser and SMF-28 fibers operated below the single-mode cutoff wavelength, we explore nonlinear mode dynamics through precise coupling, holographic mode decomposition, and spectral analysis. Experiments on 12-meter and 50-meter fibers reveal significant energy transfer to higher-order modes via intermodal Raman scattering. These findings advance our understanding of nonlinear propagation in FMFs, with important implications for high-power laser delivery and next-generation optical communication systems.
\end{abstract}

\begin{IEEEkeywords}
Few-mode fiber (FMF), Nonlinear dynamics, Intermodal Raman Scattering, Holographic Mode-decomposition, Higher-Order Modes (HOMs), Energy Transfer.
\end{IEEEkeywords}

\section{Introduction}
\IEEEPARstart{T}{he} relentless human pursuit to master light, the most essential medium for information and energy, has, for decades, manifested its most refined form within the realm of optical fibers. A delicate interplay between linear and nonlinear phenomena governs the propagation of high-intensity light pulses through these waveguides. For a long time, single-mode fiber (SMF) stood as the undisputed cornerstone of optical communications and a multitude of laser applications; its dominance was attributed to its well-defined characteristics and exceptionally low dispersion. However, the continuous march of technological advancement, coupled with an insatiable demand for ever-greater information throughput and the imperative for superior power handling, has compelled the attention of optical researchers to the realms of multimode fibers (MMFs) and few-mode fibers (FMFs) \cite{wright2022nonlinear,richardson2010high}. These fibers, capable of guiding multiple spatial modes, unveil a promising trajectory toward unprecedented spectral efficiency through mode/space division multiplexing (MDM/SDM), providing a genuine means to surpass the capacity constraints of current optical networks \cite{bai2012mode,richardson2013space}. Beyond the demands of telecommunications, the rich tapestry of their modal diversity makes them uniquely advantageous for the robust transmission of high-power beams and the exploration of the fertile landscape of complex nonlinear optical interactions \cite{renninger2013optical,wright2017spatiotemporal}.

A particularly effective strategy for investigating FMF-like behavior involves reconfiguring conventional SMFs, coaxing them to operate at wavelengths below their fundamental single-mode cutoff \cite{Gloge:71}. In this specific spectral regime, higher-order modes (HOMs), which are ordinarily confined to evanescent decay, find themselves guided, thereby transforming the SMF into a functional FMF capable of supporting a limited number of spatial modes \cite{kitayama2017few}. This change, while brilliant, creates a slew of novel complexities. The simultaneous presence of numerous guided modes causes significant intermodal dispersion, promotes subtle yet pervasive modal coupling, and triggers a range of nonlinear effects. Self-phase modulation (SPM), cross-phase modulation (XPM), and four-wave mixing (FWM) are well-documented examples; however, the phenomenon of stimulated Raman scattering (SRS) \cite{Agrawal:2019, ferreira2017semi} is of paramount interest for our present investigation. SRS, an inelastic scattering process fundamentally driven by the vibrational modes of the silica matrix itself, facilitates the transfer of optical energy from shorter to longer wavelengths—a phenomenon universally known as the Stokes shift. In the intricate confines of FMFs, this energy cascade can manifest not merely within the initially excited fundamental mode but also across several distinct spatial modes, resulting in complex and often counterintuitive patterns of spectral and spatial energy redistribution \cite{pourbeyram2013stimulated,eslami2022two}. Such intermodal Raman scattering is conveniently used for signal amplification in SDM systems \cite{ZITELLI2024103854} and can, indeed, precipitate a significant depletion of power from the fundamental mode via SRS intermodal coupling, thereby compromising both the efficiency and stability of FMF-based systems\cite{fu2023spatiotemporal,fu2025performance}.

While previous studies have rigorously documented instances of beam self-cleaning phenomena—wherein the nonlinear evolution of light within a fiber appears to preferentially favor the fundamental mode\cite{Krupa2017,guenard2017kerr}—our meticulous measurements reveal an unexpected and opposing trend. As the optical power introduced into solid-core FMFs increases, we observe a strong nonlinear modal instability, characterized by a substantial depletion of the fundamental LP$_{01}$ mode. This enigmatic behavior, we contend, originates from a subtle yet profound mechanism: the differential Raman gain experienced across the various spatial modes. This disparity, in turn, arises from the unique spatial profiles of these modes and the subtle differences in their effective refractive indices, creating a selective amplification that ultimately undermines the stability of the desired fundamental mode propagation.

\section{Experimental setup}
\begin{figure}[ht]
\centering
\includegraphics[width=1.0\linewidth]{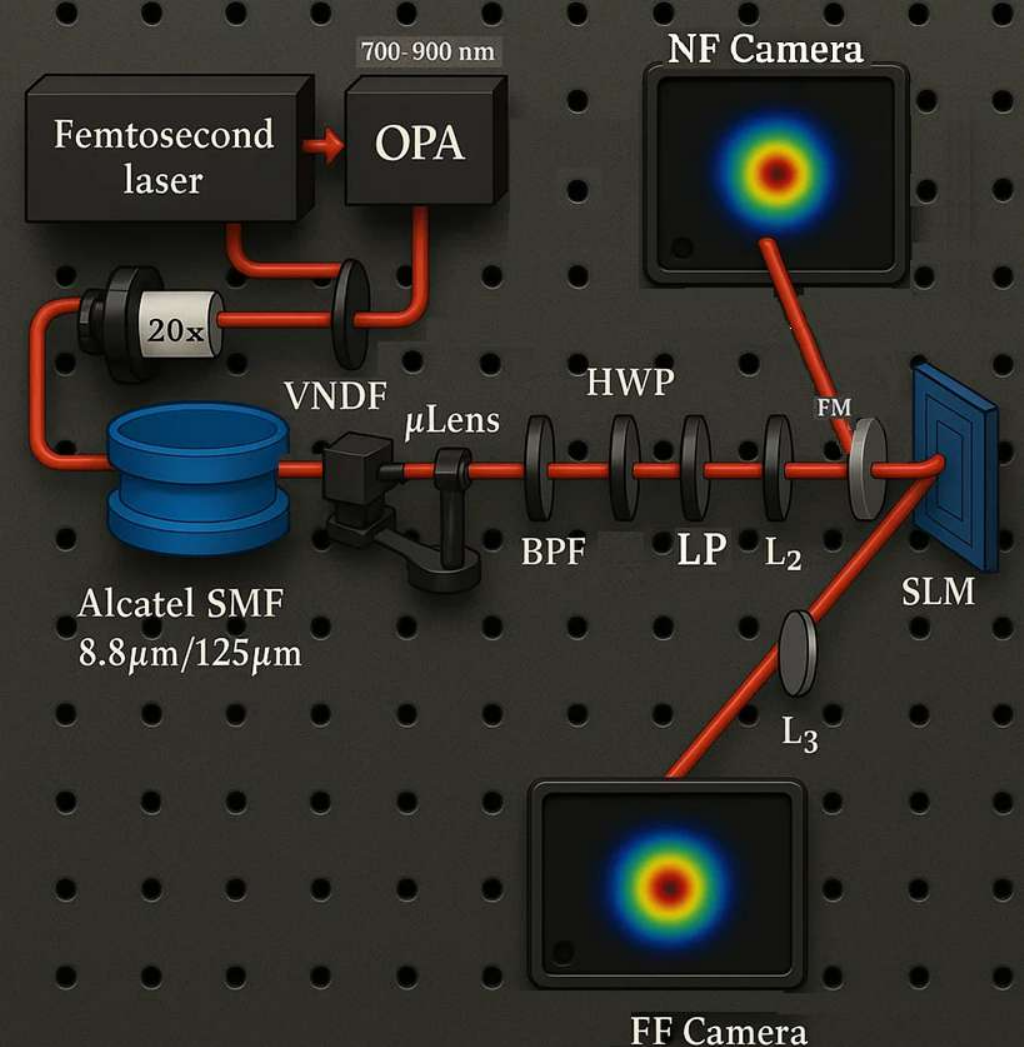}
\caption{A schematic of the experimental setup. VNDF: Variable neutral density filter; L1-L3: Lens; BPF: Bandpass filter; HWP: half-wave plate; LP: Linear polarizer; FM: Flip mirror; NF $\&$ FF camera: Near and Far field camera; SLM: Spatial light modulator.}
\label{fig:setup}
\end{figure}

Figure~\ref{fig:setup} presents the experimental setup used for this study. The laser source was a hybrid optical parametric amplifier (OPA) system (Light Conversion ORPHEUS-F),  seeded by a white-light continuum and pumped by a femtosecond ytterbium-doped laser (Light Conversion PHAROS-SP-HP). This system generates linearly polarized pulses of $70\text{ fs}$ duration at a repetition rate of $100\text{ kHz}$. The central wavelength was tunable from $650$ to $900\text{ nm}$ and fixed at 700, 800, and $900\text{ nm}$-deliberately chosen to fall below the modal cut-off wavelength of the standard SMF-28 fiber (around $1450\text{ nm}$ \cite{mangini2021experimental}), thereby ensuring the excitation of only a few spatial modes. The laser beam, possessing a near-Gaussian profile and a beam quality factor $M^2\: = \:1.3$, was precisely focused into the fiber core using a $20\times$ microscope objective (with a numerical aperture of $NA\: = \:0.4$). A variable neutral density filter (VNDF), placed just before the objective, allowed fine control of the input average power ranging from $0.1$ to $5\text{ mW}$, thereby yielding peak powers close to $100\text{ kW}$. The fiber itself, a standard Alcatel SMF-28, with a core diameter of $8.8\:\mu m$ and a cladding diameter of $125\:\mu m$, with a cladding refractive index of $n_{clad}\: =\: 1.4574$ at $630\text{ nm}$. The relative index contrast was $\Delta n \:= \: 0.003$. A subtle index dip in the center of the core is characteristic of this fiber. 

To investigate the beam propagation dynamics across varying interaction lengths, two distinct fiber lengths were used: one measuring 12 meters and the other 50 meters. The coupling into the fiber core required precise alignment via a three-axis translation stage with micrometer precision. When fastidiously aligned on-axis, the coupling efficiency—the ratio of transmitted to input power under linear conditions—averaged around $75\%$. Furthermore, approximately $\sim 4\%$ of the input power resides within $10\text{ nm}$ of the $700\text{ nm}$ central wavelength, a testament to the significant power carried by the spectral components outside this narrow band. Although our primary goal was to excite the fundamental $\text{LP}_{01}$ mode, even minor misalignment could result in contributions from HOMs.

At the fiber output end, the laser beam was collimated using a microlens ($\mu\text{Lens}$, $f = 2.7\text{ mm}$) and an uncoated lens ($\text{L}_2$, $f =400\text{ mm}$). The beam was then directed towards a Hamamatsu spatial light modulator (SLM) via a half-wave plate (HWP) and a linear polarizer (LP). These optical elements, combined, provide an exquisite means of finely tuning the beam's intensity and polarization before its interaction with the SLM, thereby increasing modulation efficiency. A flip mirror (FM) allowed for quick switching between two detection pathways: one for imaging the near-field beam profile onto a CCD camera (NF camera) and another for capturing the Fourier transform of the beam onto a second CCD (FF camera) positioned after a 500 mm convex lens ($\text{L}_3$).  We were able to precisely project the output onto specified spatial modes by adding tailored phase masks to the SLM and quantifying their power contributions, as described in Ref. \cite{gervaziev2020mode}. To eliminate the unwanted specters of self-phase modulation, which emerge as spectral broadening or temporal decoherence, a narrowband bandpass filter ($10\text{ nm}$ bandwidth) was strategically placed immediately after the fiber's exit facet, exactly centered at the pump wavelength (700, 800, or 900 nm). This ensured that only the desired spectral components were analyzed downstream. The output spectrum, spanning from 600 to 1700 nm, was then recorded by an optical spectrum analyzer (OSA) (Yokogawa AQ6370D), which allowed us to identify and exclude parasitic signals, such as Stokes and anti-Stokes shifts.

\section{Experimental result}\label{sec:ExpRes}
\begin{figure*}[!t]
\subfloat{%
    \begin{overpic}[width=0.26\textwidth]{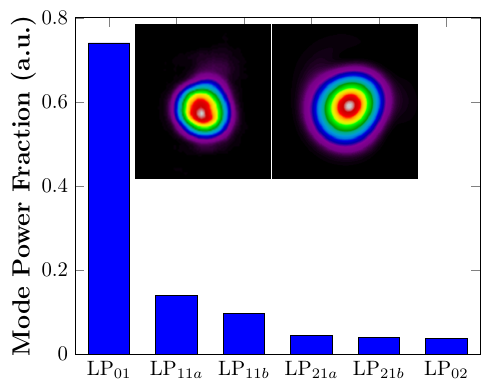} %
        \put(120,90){\bfseries\small a)} %
    \end{overpic}%
\label{fig:4a}
}
\hfil \hspace{-3.0mm}
\subfloat{%
    \begin{overpic}[width=0.245\textwidth]{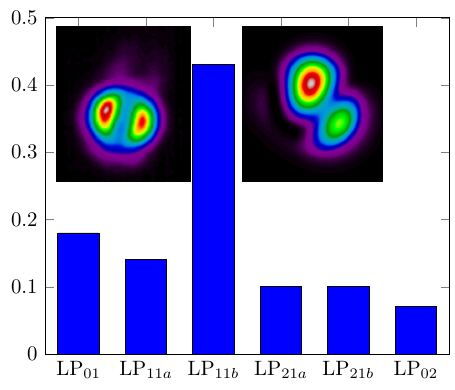}
        \put(110,90){\bfseries\small b)}
    \end{overpic}%
\label{fig:4b}
}
\hfil \hspace{-3.0mm}
\subfloat{%
    \begin{overpic}[width=0.245\textwidth]{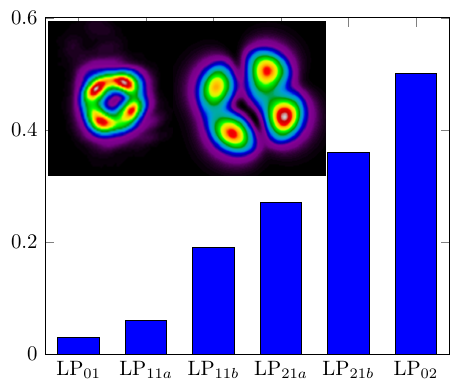}
        \put(110,92){\bfseries\small c)}
    \end{overpic}%
\label{fig:4c}}
\hfil \hspace{-1.0mm}
\subfloat{%
    \begin{overpic}[width=0.245\textwidth]{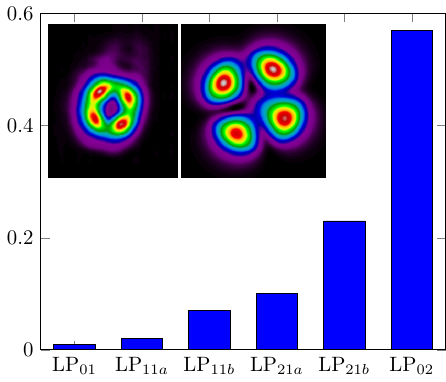}
        \put(95,90){\bfseries\small d)}
    \end{overpic}%
\label{fig:4d}}
\caption{Experimentally measured mode power distributions for fiber length of $50\text{ m}$ at center wavelength of $\lambda \:= \:700\text{ nm}$ for input average powers of a) $P_{in}\:= \:0.1\text{ mW}$, b) $P_{in}\:= \:1.6\text{ mW}$, c) $P_{in}\:= \:4.5\text{ mW}$, and d) $P_{in}\:= \:5.0\text{ mW}$.}
\label{fig:4}
\end{figure*}

The guided modes supported by the optical fiber were computed using the semi-vectorial finite difference method, based on the experimentally measured refractive index profile. This numerical approach, presented in section \ref{sec:NumRes}, demonstrates that the fiber can propagate six LP modes at a wavelength of $700\text{ nm}$: $LP_{01}$, $LP_{11a}$, $LP_{11b}$, $LP_{21a}$, $LP_{21b}$, and $LP_{02}$. To accurately resolve the modal composition at the fiber output, we employed a holographic mode decomposition technique, as comprehensively delineated in Ref. \cite{gervaziev2020mode}. The resulting power fractions associated with each LP mode are detailed in Fig.~\ref{fig:4}. The inset, in particular, shows a comparison between the experimentally obtained near-field intensity profile (left panel) and the reconstructed modal distribution generated from FF camera data (right panel), which is perfectly consistent with the methods given in Ref. \cite{gervaziev2020mode}. Under conditions of low-power linear propagation, specifically for an input power of $0.1\text{ mW}$ (corresponding to a pulse energy of $E_p=1\text{ nJ}$) with on-axis coupling, the output beam predominantly maintains the spatial profile characteristics of the fundamental mode of the fiber. As illustrated in Fig.~\ref{fig:4a}, the $\text{LP}_{01}$ mode carries most of the optical power, with only minor contributions from HOMs. However, as the input power increases, the modal composition undergoes a marked evolution. At $1.6\text{ mW}$ ($E_p=16\text{ nJ}$), the $\text{LP}_{11a/b}$ modes become significantly populated, as shown in Fig.~\ref{fig:4b}. Upon entering strongly nonlinear regime, with input powers of $4.5\text{ mW}$ ($E_p=45\text{ nJ}$) and $5.0\text{ mW}$ ($E_p=50\text{ nJ}$), the fundamental mode gets significantly depleted, and its energy is redistributed into higher-order spatial modes, including $\text{LP}_{21a/b}$ and $\text{LP}_{02}$, as compellingly illustrated in Figs.~\ref{fig:4c} and ~\ref{fig:4d}. 

\begin{figure}[!t]
\subfloat{%
    \begin{overpic}[width=0.251\textwidth]{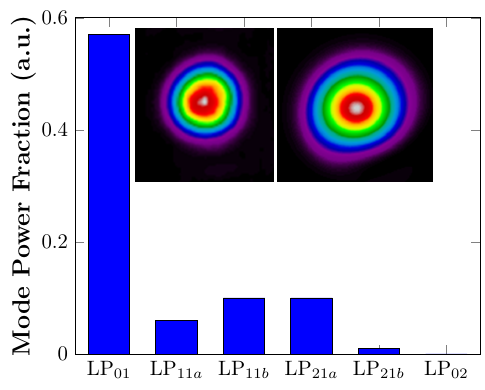} %
        \put(117,90){\bfseries\small a)} %
    \end{overpic}%
\label{fig:5a}
}
\hfil \hspace{-3.0mm}
\subfloat{%
    \begin{overpic}[width=0.237\textwidth]{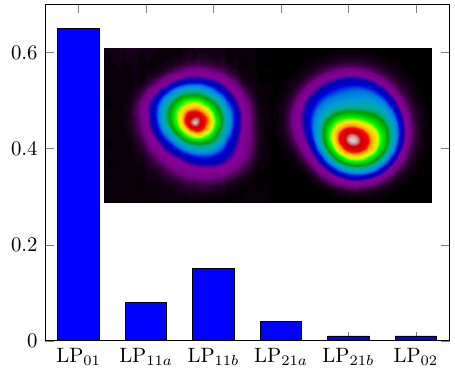}
        \put(110,90){\bfseries\small d)}
    \end{overpic}%
\label{fig:6a}
}
\vfil
\subfloat{%
    \begin{overpic}[width=0.25\textwidth]{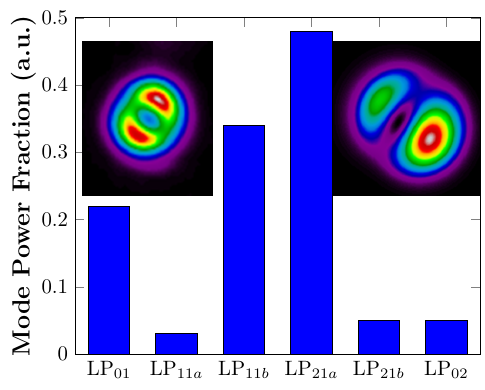} %
        \put(65,90){\bfseries\small b)} %
    \end{overpic}%
\label{fig:5b}
}
\hfil \hspace{-3.0mm}
\subfloat{%
    \begin{overpic}[width=0.237\textwidth]{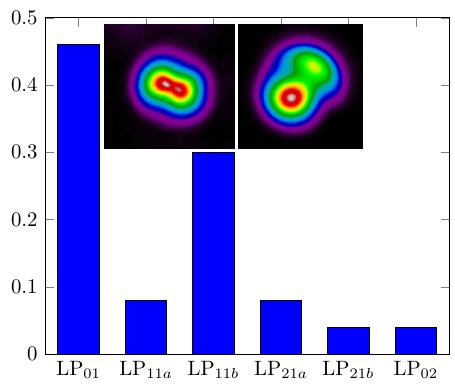}
        \put(110,90){\bfseries\small e)}
    \end{overpic}%
\label{fig:6b}
}
\vfil
\subfloat{%
    \begin{overpic}[width=0.25\textwidth]{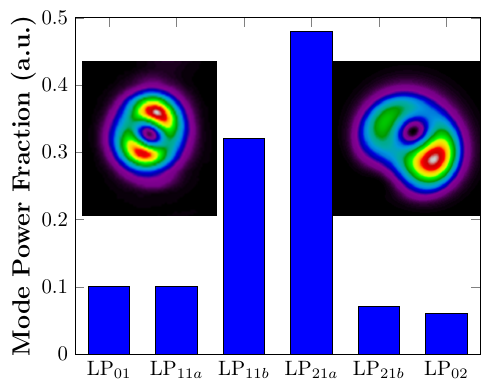} %
        \put(65,90){\bfseries\small c)} %
    \end{overpic}%
\label{fig:5c}
}
\hfil \hspace{-3.0mm}
\subfloat{%
    \begin{overpic}[width=0.237\textwidth]{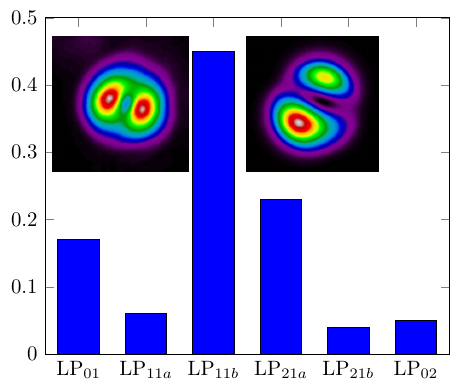}
        \put(110,90){\bfseries\small f)}
    \end{overpic}%
\label{fig:6c}
}
\caption{ Experimental mode power distributions at the fiber output for two center wavelengths and varying input powers: (a–c) At $\lambda \:= \:800\text{ nm}$: a) $P_{in}\:= \:0.1\text{ mW}$, b) $P_{in}\:= \:1.1\text{ mW}$, and c) $P_{in}\:= \:6.6\text{ mW}$. (d–f) At $\lambda \:= \:900\text{ nm}$: d) $P_{in}\:= \:0.1\text{ mW}$, e) $P_{in}\:= \:1.8\text{ mW}$, and f) $P_{in}\:= \:5.0\text{ mW}$, for 50-m-long fiber.}
\label{fig:5_6}
\end{figure}

Crucially, this transformative reconfiguration of the modal occupancy persists across distinct pump wavelengths. A strikingly similar behavior is observed for input laser carrier wavelengths of $800\text{ nm}$ and $900\text{ nm}$, as clearly illustrated in Fig.~\ref{fig:5_6}. The near-field intensity and corresponding modal decomposition at the fiber exit facet for varying input powers are showcased in Figs.~\ref{fig:5_6}(a)-(c) for $800\text{ nm}$ and Figs.~\ref{fig:5_6}(d)-(f) for $900\text{ nm}$. In the linear regime, with an input power of $0.2\text{ mW}$, the beam exhibits a quintessential bell-shaped intensity profile, a hallmark unmistakable of an output dominated by the fundamental mode. However, as the input power exceeds $1.2 \text{ mW}$, the power progressively shifts toward HOMs—most prominently $\text{LP}_{11a/b}$ and their coherent superposition-clearly signaling strong nonlinear coupling. At the highest pump levels, the output beam structure strikingly resembles that of an orbital angular momentum (OAM) mode or vortex beam, underscoring the pronounced, wavelength-sensitive dynamics of nonlinear mode excitation.

\begin{figure}[ht!]
    \centering
    \includegraphics[width=1.0\linewidth]{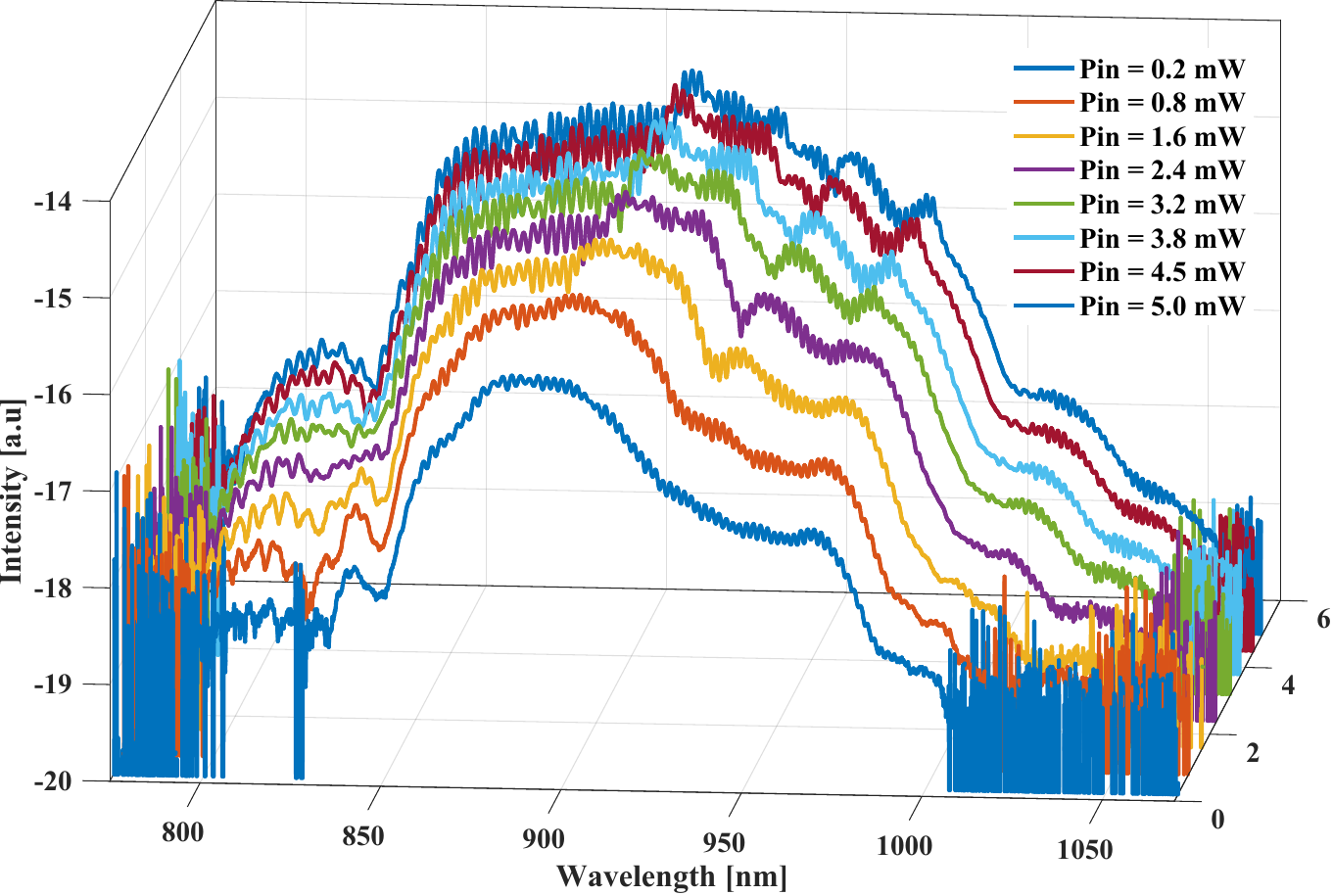}
    \caption{Experimental spectra obtained from a 50-m-long Alcatel SMF-28 as a function of the input guided power measured at the input for a laser pulse centered at $900\text{ nm}$.}
    \label{fig:6}
\end{figure}

Figure~\ref{fig:6} depicts an experimentally observed spectrum with a very broad spectrum centered at $900\text{ nm}$ and reaching around $40\text{ nm}$ in width. The spectrum shows a notable redshift and an abrupt truncation on its blue flank, providing unequivocal evidence of asymmetric spectral widening. The lack of obvious additional sidebands strongly suggests that this broadening arises from a complex and dynamic interplay between SPM and SRS, rather than from phase-matched parametric processes. This interaction is not cumulative; instead, it involves a competitive redistribution of energy within the pulse envelope. Ordinarily, SPM would orchestrate a symmetric spectral broadening, shaping both the red and blue wings of the spectrum equally. However, SRS intervenes with notable efficacy, redirecting a substantial portion of the pump pulse's energy into red-shifted Stokes components. In doing so, it significantly depletes the pulse’s peak intensity—a critical factor for efficient SPM—and consequently impedes the generation of its characteristic blue-shifted constituents. This phenomenon is particularly pronounced in the normal dispersion regime, where the Stokes wave lags the pump, thereby prolonging its interaction length and increasing energy transfer. Simultaneously, SRS induces a progressive red shift via the Raman gain, augmenting the spectral density on the red spectral tail. The resulting spectral shape—truncated in the blue and expanded in the red—is the fascinating consequence of a complicated tug-of-war in which SRS both undermines and overwhelms SPM in a delicate energy balance.   

\begin{figure}[ht!]
    \centering
    \includegraphics[width=1.0\linewidth]{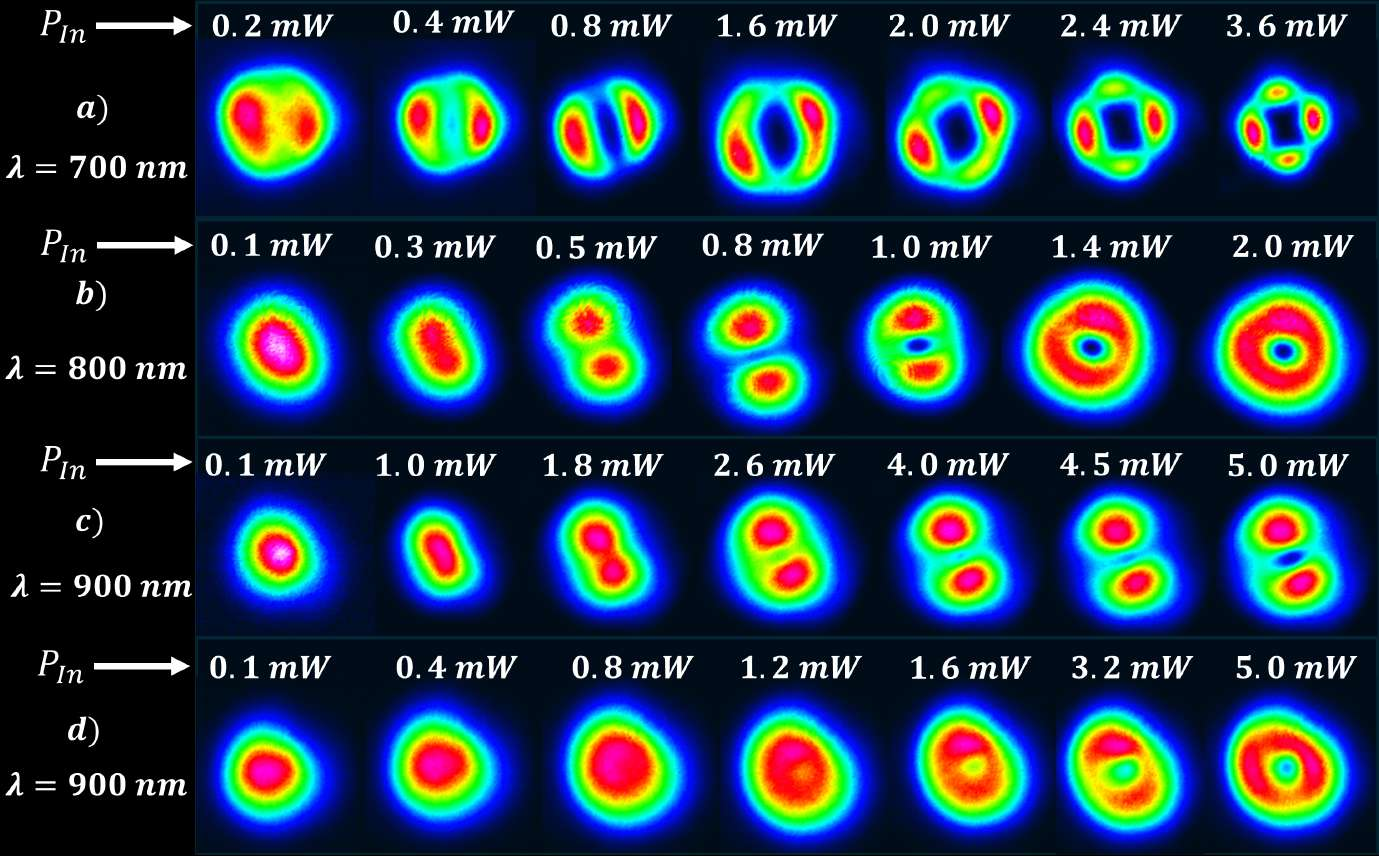}
    \caption{A fundamental mode reshaping near-field profile in a $50\text{ m}$-long fiber with input wavelengths of (a) $700\text{ nm}$, (b) $800\text{ nm}$, and (c) $900\text{ nm}$. (d) With an offset input, the same as (c), for different input average power $\text{P}_{\text{In}}$. }
    \label{fig:7}
\end{figure}

To further investigate the spatial dynamics that accompany this spectrum evolution, we carefully removed the band-pass filter at the fiber's output and recorded a series of near-field intensity profiles corresponding to input pulses at $700\text{ nm}$, $800\text{ nm}$, and $900\text{ nm}$. These illuminating tests unequivocally indicated a significantly rapid depletion of the fundamental mode at relatively low input levels, owing once again to the presence of the Raman effect. Fig.~\ref{fig:7}a vividly illustrates how the output beam transitions from a $\text{LP}_{11}$-like mode to an $\text{LP}_{21}$-like structure with noticeable center power depletion, a configuration distinct from the mode structure observed in Fig.~\ref{fig:4}. This intriguing evolution is further corroborated at longer wavelengths. Fig.~\ref{fig:7}b demonstrates that under $800\text{ nm}$ excitation and low input power, the output beam exhibits a near-Gaussian profile, indicative of the dominance of the fundamental mode. When the input power exceeds $\text{P}_{In}\:=\:1.4\text{ mW}$, the beam undergoes a remarkable metamorphosis from an $\text{LP}_{21}$-like profile to a structure resembling an optical vortex (OAM-like mode). Surprisingly, this vortex-like phase remains stable even when subjected to further power increases or external perturbations. A nearly similar progression occurs at $900\text{ nm}$.  Figure~\ref{fig:7}c captures how a bell-shaped fundamental mode produces a higher-order $\text{LP}_{21}$-like profile with a hollow center. Intriguingly, this transition has surprising robustness: regardless of the initial coupling conditions—be it axial alignment or a deliberate lateral offset—the beam always moves towards an OAM-like shape as the input power approaches $5\text{ mW}$. This profound universality suggests an underlying nonlinear mode-selection mechanism, intricately regulated by the concerted interaction of SPM, SRS, and modal dispersion.

\begin{figure*}[!t]
\subfloat{%
    \begin{overpic}[width=0.26\textwidth]{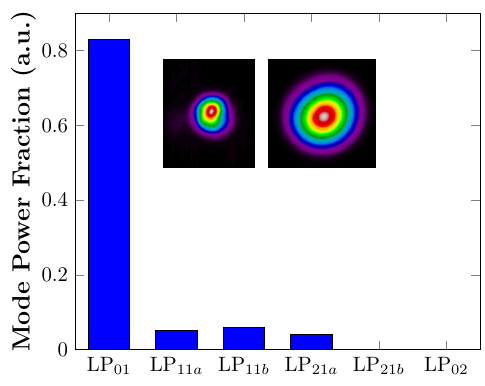}%
        \put(120,92){\bfseries\small a)} %
    \end{overpic}%
\label{fig:8a}
}
\hfil \hspace{-3.0mm}
\subfloat{%
    \begin{overpic}[width=0.245\textwidth]{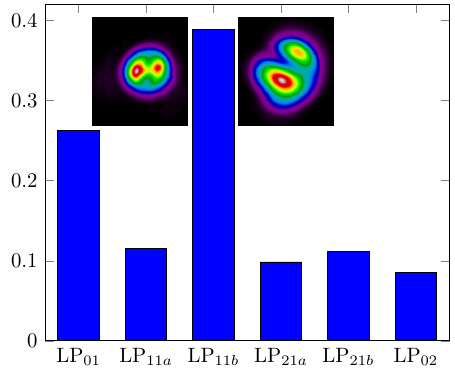}
        \put(110,92){\bfseries\small b)}
    \end{overpic}%
\label{fig:8b}
}
\hfil \hspace{-3.0mm}
\subfloat{%
    \begin{overpic}[width=0.245\textwidth]{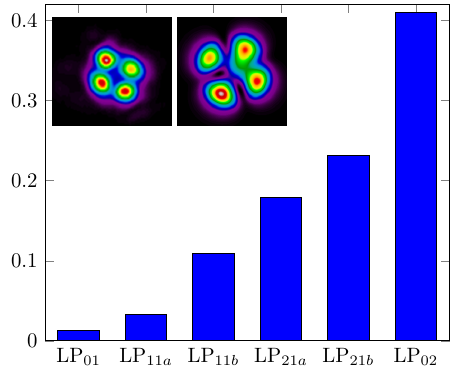}
        \put(90,92){\bfseries\small c)}
    \end{overpic}%
\label{fig:8c}}
\hfil \hspace{-1.0mm}
\subfloat{%
    \begin{overpic}[width=0.245\textwidth]{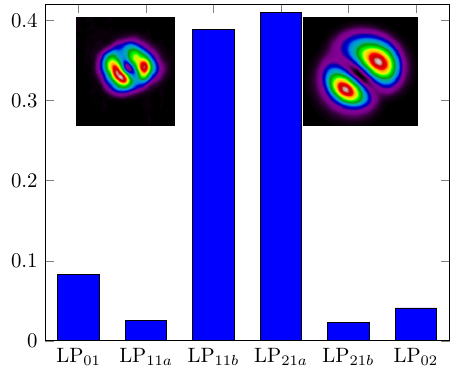}
        \put(117,92){\bfseries\small d)}
    \end{overpic}%
\label{fig:8d}}
\caption{Experimentally measured mode power distributions for fiber length of $12\text{ m}$ at center wavelength of $\lambda \:= \:700\text{ nm}$ for input average powers of a) $P_{in}\:= \:0.1\text{ mW}$, b) $P_{in}\:= \:2.0\text{ mW}$, c) $P_{in}\:= \:4.0\text{ mW}$, and d) $P_{in}\:= \:6.0\text{ mW}$.}
\label{fig:8}
\end{figure*}

To deepen our understanding of how fiber length influences Raman-driven modal power transfer, a complementary experiment was conducted using a $12\text{ m}$-long fiber, which is shorter than in previous experiments. This fiber was excited with a bell-shaped beam centered at $700\text{ nm}$. Fig.~\ref{fig:8} presents the evolution of the modal power distribution at the output, measured for varying input power levels. At a lowest input of $\text{P}_{In} = 0.1\text{ mW}$ (Fig.~\ref{fig:8}(a)), the output remains overwhelmingly confined to the fundamental $\text{LP}_{01}$ mode. This dominance reflects predominantly linear propagation, with negligible energy transfer to HOMs. As the input increases to $\text{P}_{In} = 2.0\text{ mW}$ (Fig.~\ref{fig:8}(b)), the first signs of nonlinear behavior emerge. While the $\text{LP}_{01}$ mode still carries a substantial portion of the power, a noticeable shift occurs: the energy is gradually diverted into $\text{LP}_{11a}$ and $\text{LP}_{11b}$, and to a lesser extent, $ \text{LP}{21a}$, $\text{LP}{21b}$, and $\text{LP}_{02}$. This signals the beginning of intermodal Raman interactions. Further increasing the input to $\text{P}_{In} = 4.0\text{ mW}$ (Fig.~\ref{fig:8}(c)) enhances this redistribution. The $\text{LP}_{01}$ mode gradually fades, replaced by more noticeable contributions from the higher-order $ \text{LP}{21a/b} $. At the highest input of $\text{P}_{In} = 6.0\text{ mW}$ (Fig.~\ref{fig:8}(d)), the modal landscape is transformed. The $\text{LP}_{11a/b}$ modes now dominate the output, with significant power also coming from the $\text{LP}_{21a/b}$. The fundamental $\text{LP}_{01}$ mode is significantly reduced, indicating an effective transfer of energy to higher-order spatial modes due to the Raman effect.

\section{Numerical results}\label{sec:NumRes}
The propagation of ultrashort pulses in an Alcatel SMF-28 optical fiber operating below its cutoff wavelength, which supports $M$ different spatial modes, can be accurately described by the generalized multimode nonlinear Schrödinger equation (GMMNLSE) \cite{Poletti:08,Wright2018a}. This model captures the complex spatiotemporal evolution of the optical field in a few-mode fiber (FMF). The total optical field envelope is expressed as a superposition of the fiber's eigenmodes:
\begin{equation}
\small
    E(x,y,z,t) = \sum_{p=1}^M F_p(x,y) A_p(z,t)
\end{equation}
where $F_p(x,y)$ denote the transverse mode profiles and $A_p (z,t)$ represents a slowly varying complex temporal envelope associated with the $p^{th}$ mode. For the purpose of this study, considering $M=6$ modes, the GMMNLSE for the envelope of the $p^{th}$ mode, $A_p(z,t)$, is given by:

\begin{align}
\small
\begin{split}
&\partial_z A_p (z,t) = i\left( \beta^p_0-\beta_0 \right) A_p - \left( \beta^p_1-\beta_1 \right) \frac{\partial A_p}{\partial t} +  \sum^4_{q\geq 2} i^{q+1} \frac{\beta_q^p}{q!} \partial_t^m A_p  \\
& - \frac{\alpha_p}{2} A_p + i \frac{n_2 \omega_0}{c} \left( 1+\frac{i}{\omega_0} \partial_t \right) \sum_{l,m,n} \{ (1-f_R) S^k_{plmn} A_l A_m A_n^* \\
& + f_R A_l S^R_{plmn} \int^t_{-\infty} A_m(z,t-\tau) A_n^*(z,t-\tau) h_R(\tau) d\tau\}
\end{split}
\label{eq:GMMNLSE}
\end{align}
The coefficient $\beta_q^{(p)}$ represents the $q^{th}$ order dispersion term for the $p^{th}$ mode. The first two terms account for the phase mismatch and the group velocity mismatch among modes. The third term describes higher-order chromatic dispersion. Nonlinear contributions—mediated by the Kerr coefficient $n_2=2.7\times 10^{-20} m^2/W$, central frequency $\omega_0$, and nonlinear overlap integrals $S^k_{plmn}$—capture instantaneous nonlinear effects (SPM, XPM, FWM). The Raman effect enters via the fractional response $f_R\approx 0.18$, Raman overlap integrals $S^R_{plmn}$, and a delayed response function $h_R(\tau)$ parameterized by $\tau_1 = 12.2\text{ fs}$ and $\tau_2=32\text{ fs}$. Modal specific attenuation is accounted for by loss coefficients $\alpha_p$.

As detailed in the experimental section, the simulation employed an Alcatel SMF-28 with a step-index core diameter of $8.8\: \mu m$ and an index contrast of $\Delta n = 0.003$ between its peak core and surrounding cladding. At a wavelength of $700\text{ nm}$, the fiber exhibited normal chromatic dispersion, with $\beta_2$ ranging from $22.2$ to $34.9\: ps^2/km$ across $LP_{01}$ to $LP_{02}$ modes, respectively. Furthermore, the modal delays relative to the fundamental $LP_{01}$ mode were 2.0, 2.0, -3.2, -3.2, and -9.3 ps per meter of fiber for modes 2 through 6. Eq.~\ref{eq:GMMNLSE} was numerically solved with an integration step of $50 \:\mu\text{m}$, utilizing a spatial window of $125\:\mu m \times 125\:\mu m$ and a transverse grid of $800\times800$ samples. An ultrashort pulse, with a duration $T_0 = 70\text{ fs}$ at a full width half maximum (FWHM) and a diameter of $2w_0 = 15\:\mu\text{m}$, was launched primarily into the fundamental mode of the FMF. The simulation explored input pulse energies varying from $1\text{ nJ}$ to $30\text{ nJ}$, across fiber lengths of $1\text{ m}$, $6\text{ m}$, and $12\text{ m}$. The initial fractional mode power distribution is depicted in Fig.~\ref{fig:4a}.
\begin{figure}[ht!]
\centering
\includegraphics[width=0.5\textwidth]{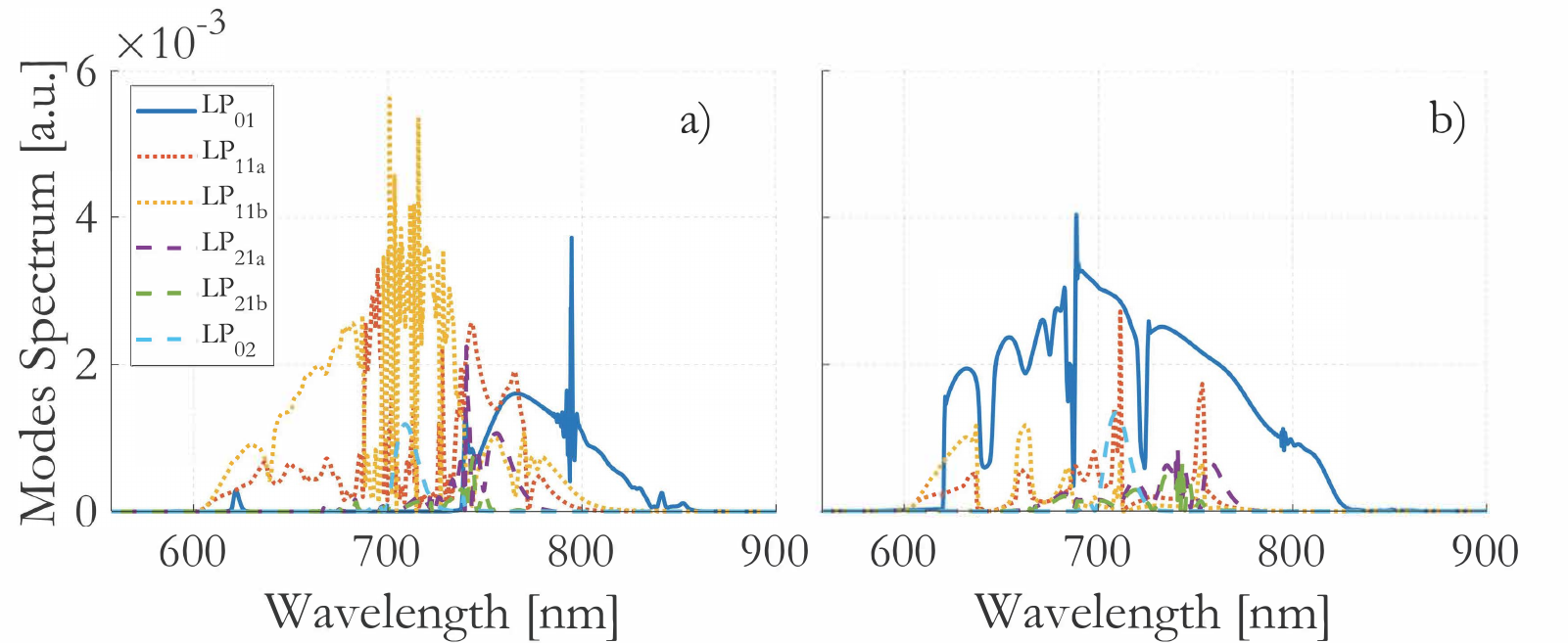}
\caption{Raman's transformative role in FMF dynamics. Comparative numerical simulations illustrate the output modal spectral profiles after 6 m of FMF, (a) with and (b) without the Raman nonlinearity. Input pulse with 70 fs pulsewidth and 30 nJ energy.}
\label{fig:9}
\end{figure}

\begin{figure}
\centering
\includegraphics[width=0.5\textwidth]{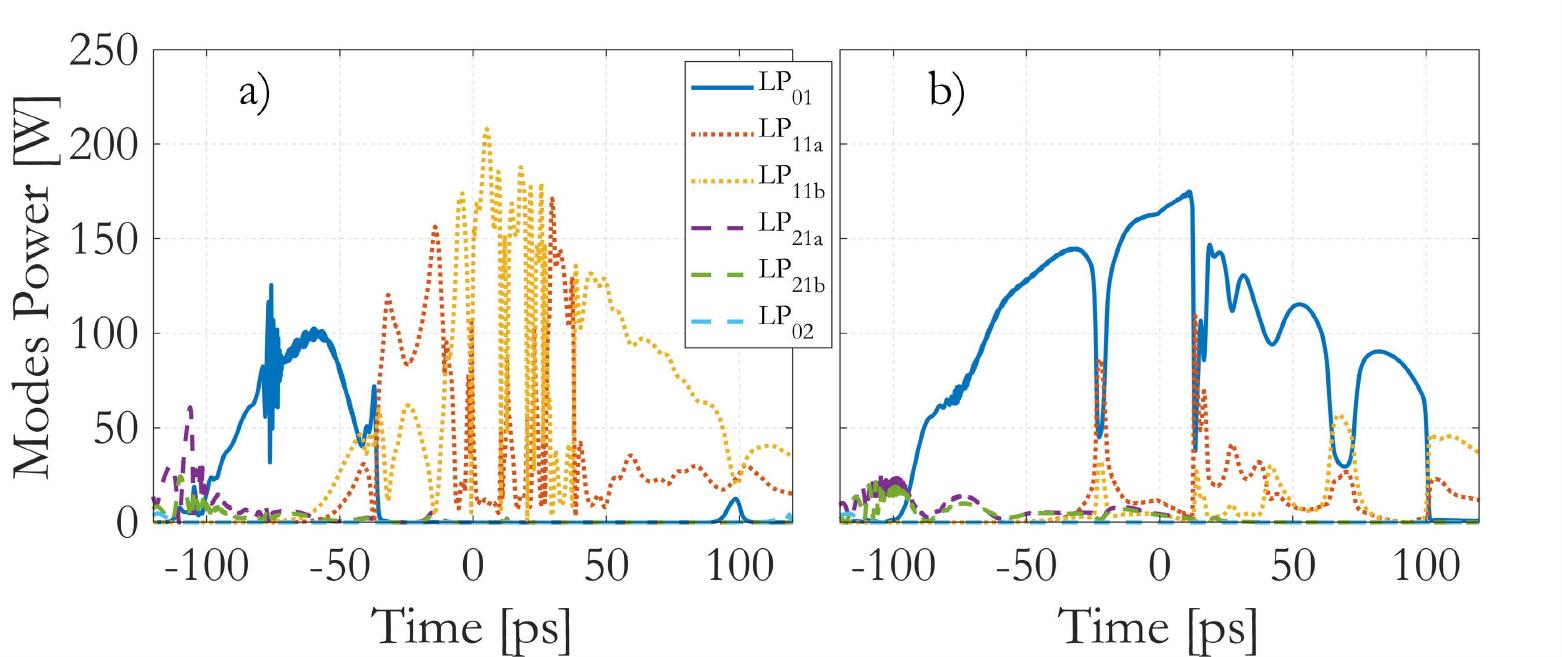}
\caption{Simulated output modal power after 6 m of FMF: (a) with, and (b) without, the Raman nonlinearity. Input pulse with 70 fs pulsewidth and 30 nJ energy.}
\label{fig:10}
\end{figure}
Figures~\ref {fig:9} and \ref{fig:10} reveal a fascinating set of numerical results, clearly demonstrating the profound impact of the Raman effect on ultrashort pulse propagation within FMFs. Consider a fiber length of 6 m and an input pulse energy of 30 nJ. In the absence of the Raman effect ($f_R =0$), the temporal profiles of the modes evolve with a serene smoothness, as evident in Fig.~\ref{fig:10}(b). Here, modal pulses undergo broadening and distortion, a direct consequence of the interplay between SPM and normal chromatic dispersion. This very interplay also orchestrates a moderate blue-shift of the second modal group (modes $LP_{11a}$ and $LP_{11b}$) and a red-shift of the third modal group (modes $LP_{21a}$, $LP_{21b}$ and $LP_{02}$) relative to the fundamental mode (Fig.~\ref{fig:9}(b)). Modal dispersion cannot temporally distinguish modal pulses due to pulse broadening; yet, XPM induces minimal nonlinear power exchange among modes. However, when Raman nonlinearity is introduced ($f_R =0.18$), SRS emerges as the dominant effect, forcing a net energy transfer from the fundamental mode to the second mode group. This results in amplification of the spectra for modes $LP_{11a}$, $LP_{11b}$, which concurrently leads to a depleted spectrum at lower wavelengths for mode $LP_{01}$, as illustrated in Fig.~\ref{fig:9}(a). In the temporal domain, the instantaneous power of the fundamental mode experiences a marked depletion in correspondence with its trailing tail, precisely where the modes of group 2, delayed by the modal dispersion, experience a Raman gain (Fig.~\ref{fig:10}(a)). At sufficiently high powers and extended interaction lengths, the onset of cascaded Raman scattering was observed. Here, the first Stokes-shifted light acts as a pump for subsequent Raman shifts, resulting in multiple distinct spectral peaks. The intermodal spectrum transfer is governed by the interaction of Raman gain and phase-matching conditions among multiple modes at different frequencies, producing a complex signature of spatiotemporal dynamics. XPM, though present, plays a minor role in comparison.

\begin{figure}[ht]
\centering
\includegraphics[width=0.5\textwidth]{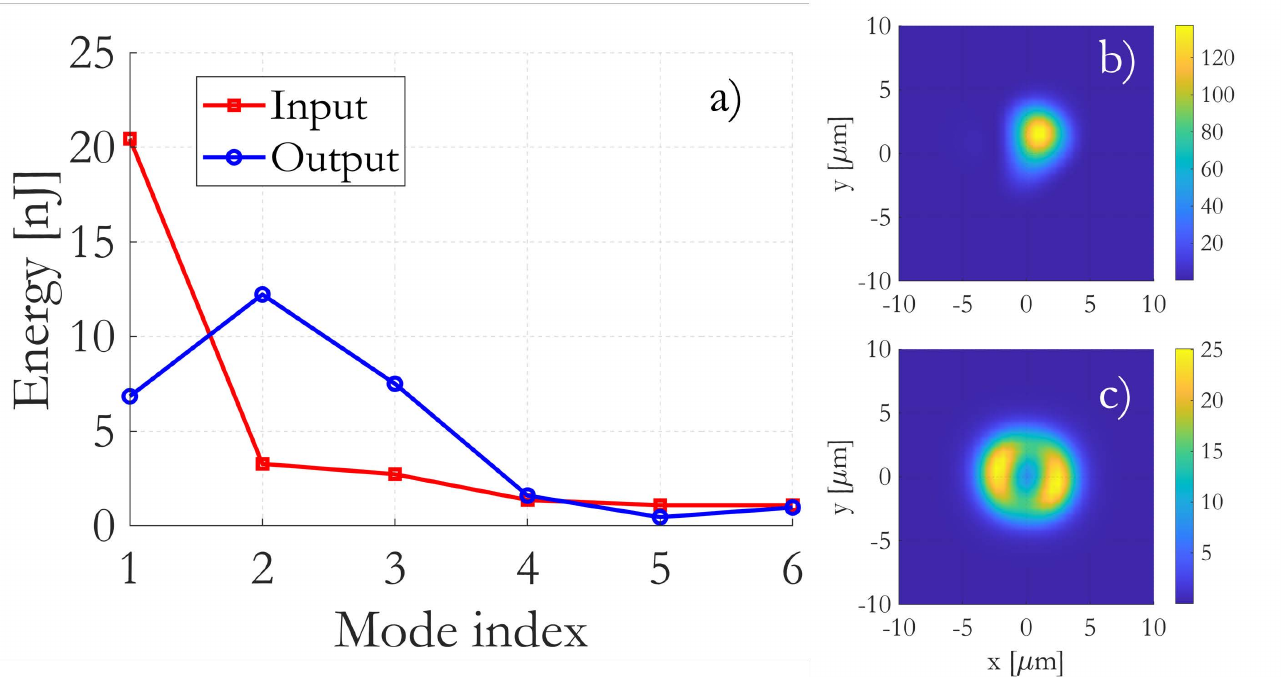}
\caption{Simulated: (a) modal energy distribution at input and after 6 m of FMF, when the input pulsewidth is 70 fs and 30 nJ total energy; (b) input near-field; (c) output near-field.}
\label{fig:11}
\end{figure}

Figure~\ref{fig:11}(a) offers compelling evidence of the simulated modal energy distribution of input and output when a 70-fs pulse, with 30 nJ energy, propagated over 6 m of FMF. In the absence of the Raman effect (not shown), the output energy in each mode remains consistently aligned with the input throughout the propagation length, exhibiting no significant transfer of power between different spatial modes during propagation. On the contrary, when the Raman effect is taken into account, the modal energies undergo dramatic and dynamic metamorphoses. The initially populated fundamental mode (mode index 1, depicted by the red curve) experiences a significant decrease in its energy as light propagates. Concurrently, the HOMs (e.g., mode group 2, represented by mode indexes 2 and 3 in the blue curve) exhibit a noticeable increase in their energy. This unmistakable observation is the direct manifestation of Raman-induced inter-modal energy transfer: energy from the depleted fundamental mode (and potentially other pump modes) is efficiently transferred to these Stokes-shifted HOMs. Thus, while the near field is dominated by the fundamental mode at input or in the absence of SRS (Fig.~\ref{fig:11}(b)), the metamorphic power of Raman nonlinearity produce an output near field that is undeniably dominated by modes $LP_{11a}$, $LP_{11b}$ (Fig.~\ref{fig:11}(c)), with the contribution of the fundamental mode rendered negligible. The efficiency and specific target modes of this energy transfer are critically dependent on the group velocity matching between the pump and Stokes components in different modes, as well as the spatial overlap integrals of the interacting modes.

In close accordance with our experimental findings, we performed a supplementary computational analysis to investigate the influence of fiber length on Raman-driven intermodal energy transfer. Simulations were carried out for two specific lengths—1 meter and 12 meters—at input pulse energies of 20 nJ and 30 nJ. In the shorter fiber, energy initially limited to the fundamental mode saw significant redistribution, with HOMs, especially the second mode group, acquiring a substantial share of the energy. The near-field intensity profile matched the simulation results given in Fig.~\ref{fig:11}(c), capturing the key aspects of modal reshaping. At 12 meters, the overall structure in the near-field remained relatively unchanged, but the energy transfer increased significantly in the third mode group. These findings reveal a surprising length independence of the nonlinear mode coupling, with Raman interactions resulting in identical behavior of energy transfer from the fundamental mode $LP_{01}$ to the HOMs with indexes of 2 ($LP_{11a}$) and 3 ($LP_{11b}$), as well as a near-field profile similar to that seen in Fig.~\ref{fig:11}(c).

\begin{figure*}[!t]
\centering
\subfloat[]{\includegraphics[width=0.4\textwidth]{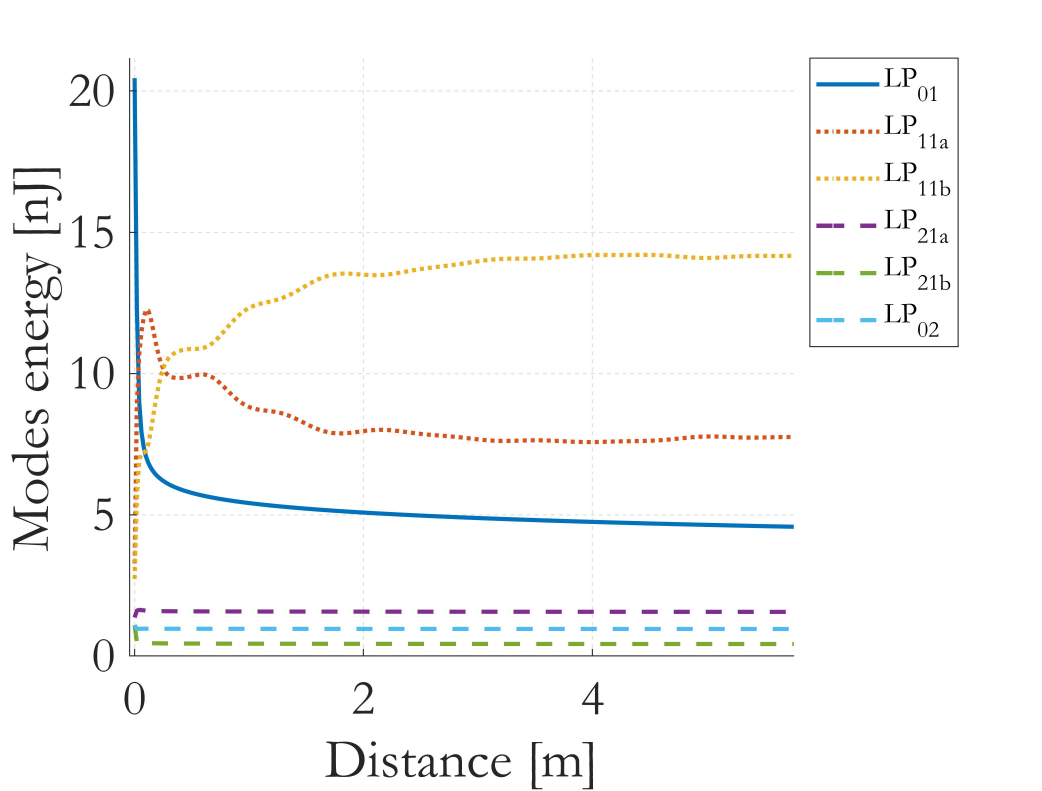}%
\label{fig:12}}
\hfil
\subfloat[]{\includegraphics[width=0.4\textwidth]{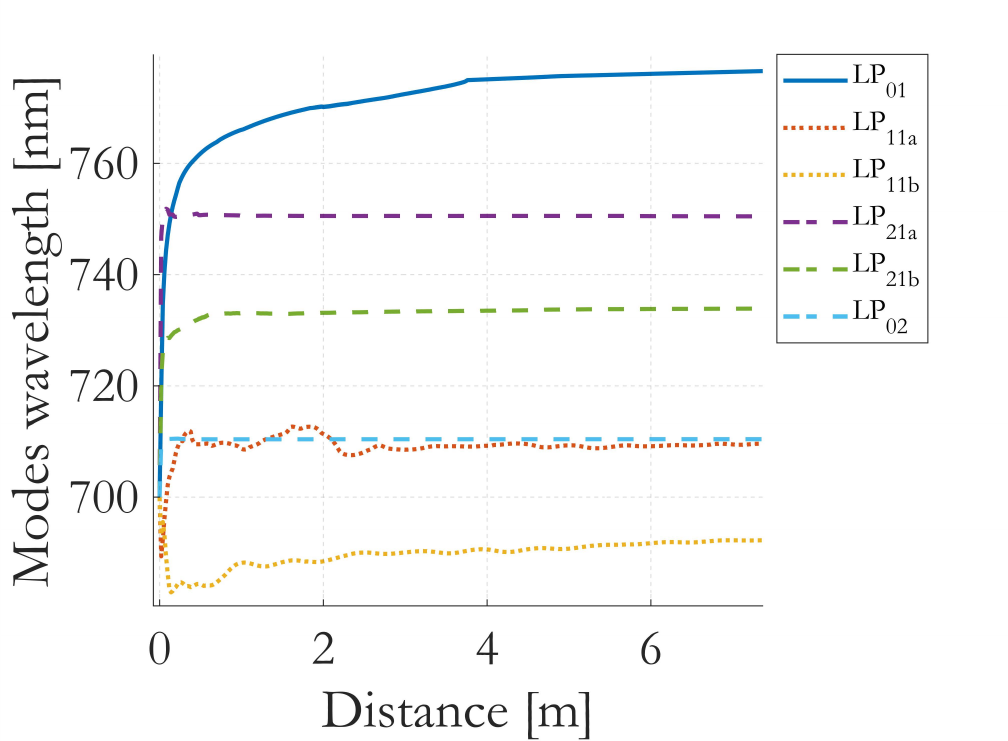}%
\label{fig:13}}
\caption{(a) Simulated evolution of modal energy and (b) simulated evolution of the central wavelength in the first 6 m of the FMF, when the input pulse width is 70 fs and the total energy is 30 nJ.}
\label{fig:12-13}
\end{figure*}

Figure \ref{fig:12} illustrates the evolution of modal energy within the first 6 meters of the FMF, following the launch of a 70-fs pulse at a wavelength of 700 nm with an input energy of 30 nJ, distributed as depicted in Fig. \ref{fig:11}. The simulation reveals a rapid migration of energy from the fundamental $LP_{01}$ mode to the $LP_{11}$ mode group. This preferential transfer is primarily due to their significant modal overlap, which facilitates an efficient exchange of optical power. In contrast, the higher-order $LP_{21}$ and $LP_{02}$ modes within group 3 show a markedly reduced energy coupling with the $LP_{01}$ mode. This intriguing dynamic ultimately stabilizes at a steady state, influenced by the temporal broadening and distortion of the pulse caused by SPM.

Turning to the frequency domain, the spectral evolution of each mode unveils a more nuanced narrative (Fig. \ref{fig:13}). While the Raman effect is universally acknowledged as a primary driver of intensity-dependent wavelength redshift, our findings introduce a captivating subtlety: the degree of redshift is not uniformly dictated by intensity across all modes. The $LP_{01}$ mode, initially endowed with the lion's share of energy, undergoes the most significant redshift, a direct consequence of strong intramodal Raman scattering. Yet, a curious paradox emerges with the $LP_{11}$ modes: despite siphoning off a substantial portion of the $LP_{01}$ mode's energy, they exhibit the least pronounced redshift. This seemingly counterintuitive behavior originates from the very nature of intermodal energy transfer. The Raman intermodal coupling between the $LP_{01}$ and $LP_{11}$ modes predominantly occurs at the shorter-wavelength components of the $LP_{01}$ mode. Consequently, the energy transferred to the $LP_{11}$ modes is predominantly situated within their short-wavelength spectral regions. This spectral localization effectively hinders the characteristic Raman redshift of the $LP_{11}$ modes, as the very process of energy acquisition acts as a counterbalance to it. In stark contrast, mode group 3, characterized by its meager overlap and weak coupling with the $LP_{01}$ mode, remains largely unaffected by this compensatory mechanism, thus preserving its inherent Raman redshift.

\section{Conclusion}\label{sec:Concl}
Our combined experimental and numerical study demonstrates that nonlinear propagation in few-mode fibers is governed by a complex interplay of SPM, SRS, and modal dispersion, resulting in dramatic changes to both spatial and spectral beam characteristics. As input power increases, energy is progressively transferred from the fundamental $LP_{01}$ mode to HOMs such as $LP_{11a/b}$ and $LP_{21a/2}$, with the emergence of vortex-like (OAM) profiles indicating robust nonlinear mode coupling. Spectral measurements confirm that the asymmetric broadening, dominated by SRS, is particularly pronounced at longer wavelengths and higher powers, with Raman-induced redshifts effectively suppressing the blue components typically generated by SPM. Simulations validate these findings, revealing that Raman interactions induce strong intermodal energy transfer and significantly reshape the modal power distribution. Notably, cascaded Raman scattering further enriches the output spectrum, highlighting the sensitivity of pulse dynamics to both input conditions and fiber properties. These results not only elucidate the nonlinear mechanisms at play in multimode fibers but also offer valuable pathways for tailoring ultrafast light fields in advanced optical systems.

\section*{Acknowledgments}
We acknowledge Vincent Couderc for providing optical fiber characterization, M. Gervaziev, D. Kharenko, and S. Babin for
providing the mode decomposition setup.

Project ECS 0000024 Rome Technopole, Funded by the European Union – NextGenerationEU.

\vfill

\end{document}